\documentclass[12pt]{article} 

\usepackage{scicite}

\usepackage{times}

\usepackage{amsmath}
\usepackage{amsfonts}
\usepackage{amssymb}
\usepackage{graphicx}
\usepackage{subcaption}
\usepackage{booktabs}

\newenvironment{sciabstract}{%
\begin{quote} \bf}
{\end{quote}}

\usepackage[hidelinks]{hyperref}
\hypersetup{allcolors=blueish,colorlinks=true}

\usepackage[labelfont=bf,figurename=Fig.,labelsep=period,font=footnotesize]{caption}

\title{Misinformation Dissemination: Effects of Network Density in Segregated Communities} 
\author
{Soroush Karimi,${}^{\ast}$ Marcos Oliveira, and Diogo Pacheco\\
\\
\normalsize{Department of Computer Science, University of Exeter, UK}\\
\normalsize{$^\ast$\href{sk931@exeter.ac.uk}{sk931@exeter.ac.uk}}
}

\date{}

\usepackage{changepage}

\usepackage{xcolor}
\usepackage{soul}
\definecolor{reddish}{HTML}{FBB4AE}
\definecolor{blueish}{HTML}{1F54A9}
\definecolor{magentish}{HTML}{FF00AA}
\definecolor{greenish}{HTML}{a1d99b}

\usepackage{hyperref}

\usepackage{setspace}

\usepackage{lineno}

\usepackage{lipsum}

\begin{document} 




\maketitle 




\begin{sciabstract}
Abstract. 

\textnormal{%
Understanding the relationship between network features and misinformation propagation is crucial for mitigating the spread of false information. Here, we investigate how network density and segregation affect the dissemination of misinformation using a susceptible-infectious-recovered framework. We find that a higher density consistently increases the proportion of misinformation believers. In segregated networks, our results reveal that minorities affect the majority: denser minority groups increase the number of believers in the majority, demonstrating how the structure of a segregated minority can influence misinformation dynamics within the majority group.
}
\end{sciabstract}

\baselineskip1.55em 

\section*{Introduction}
Misinformation poses a significant challenge in today's digitized society~\cite{muhammed2022disaster,ferrara2020characterizing,pacheco2024bots}, with social media platforms often serving as primary vectors for its dissemination, fueling polarisation~\cite{au2022role}, anti-vax sentiments~\cite{benoit2021anti}, violence~\cite{paul2023mathematical,pacheco2020unveiling}, and political interference~\cite{swire2017processing,jerit2020political}. Despite its widespread effects, our understanding of the mechanisms underlying misinformation remains elusive~\cite{meng2023impact}. In particular, we still do not fully understand the role of social groups (and their structure) in the spread of misinformation. In this work, we investigate the dynamics of misinformation spread, focusing on the role of network density in diverse social structures.

Previous works have highlighted the main role of network structure in amplifying falsehoods and the need for a better understanding of group density~\cite{chen2021covid,chen2021neutral,jost2018ideological}. For example, a higher density of social networks is associated with higher levels of dissemination of misinformation, particularly within conservative clusters~\cite{chen2021covid}. Similarly, conservative Twitter users often inhabit denser networks and are more exposed to low-credibility content, perpetuating the circulation of partisan content~\cite{chen2021neutral}. Such homogeneous social networks foster an environment conducive to misinformation proliferation, making individuals more susceptible to misinformation and triggering cascade effects that extend to the broader population~\cite{jost2018ideological}.

To understand these dynamics, simulation provides a valuable framework to gain insights into real-world phenomena and pinpoint influential factors~\cite{raponi2022fake}. For instance, Tambuscio et al. proposed a model that simulates individuals transitioning among susceptible, fact-checker, and believer states, shedding light on the mechanisms driving misinformation propagation and guiding counter-strategies~\cite{tambuscio2015fact}. However, their model focuses primarily on individual traits, such as gullibility, without accounting for the impact of network density~\cite{tambuscio2018network}.

In our work, we highlight the fundamental role of network density in both segregated and non-segregated networks. We explore how denser networks correlate with a higher percentage of believers in hoaxes and investigate the impact of a dense minority on the belief percentage within the majority.
Our results unveils a positive correlation between network density and the prevalence of \emph{believers}---individuals propagating misinformation---within a network. 
This correlation echoes findings from prior empirical studies~\cite{shao2018anatomy,del2016spreading,chen2021neutral} and suggests that denser networks facilitate misinformation dissemination. 
This trend persists within segregated networks, where increased density within groups corresponds to a greater number of believers within those groups.

We also explore cross-group effects and reveal that in networks with both majority and minority groups, the majority influences the minority, as expected. Surprisingly, however, we uncover a reciprocal effect: changing the density of the minority group---while keeping the majority's density constant---affects the majority as well.

These insights carry significant implications for mitigating and addressing misinformation~\cite{van2022misinformation}, emphasizing the critical role of our social connections. 
They demonstrate that actions from even highly segregated groups can influence the broader population, shaping collective behaviour. The interplay between these findings and the impact of superspreaders remains an open question~\cite{daverna2024superspreaders}.

\section*{Modelling misinformation spread}
To understand how network density affects the spread of the misinformation, we use a susceptible-infectious-recovered model~\cite{tambuscio2015fact}. We apply it to networks with varying densities in segregated and unsegregated scenarios.

\subsection*{The susceptible-believer-fact-checker model}

\begin{figure}[b!]
        \centering
        \includegraphics[width=0.6\textwidth]{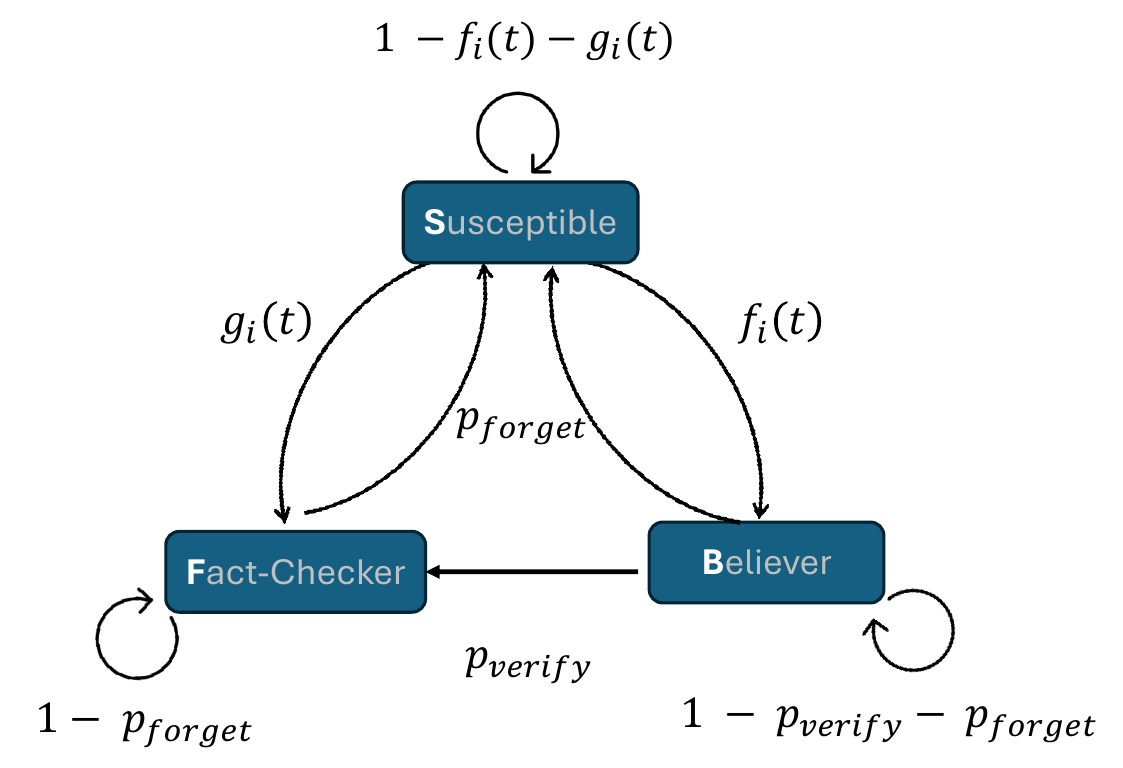}
        \caption{SBFC model -- transitions between states. Adapted from \cite{tambuscio2015fact}.}
        \label{fig:tambuscio_model}
\end{figure}

We use the susceptible-believer-fact-checker (SBFC) spread model, previously proposed for modelling misinformation dissemination in networks~\cite{tambuscio2015fact}. This model, rooted in epidemic modelling, treats hoaxes as viral infections and integrates a mechanism for fact-checking to combat their propagation. The dynamics of the model are governed by the interactions between individuals with their neighbours, each of whom can be in one of the model's states (see Fig.~\ref{fig:tambuscio_model}).

In the model, individuals in the network can be in one of three states: susceptible (S), believer (B), or fact-checker (F). Those in the susceptible state have not been exposed to the hoax or have forgotten about it. Believers have been exposed to the hoax and believe it. Finally, fact-checkers have verified the hoax and no longer believe it.

The model is characterized by four parameters:
the overall spreading rate $\beta$, the agents' gullibility $\alpha$, the probability $p_{verify}$ to verify a hoax (i.e., the probability that a believer  will fact-check a hoax and become a fact-checker), and the probability $p_{forget}$ that an individual will forget their current belief state and become susceptible again. The spread of a hoax and its debunking can be described as follows:

\begin{enumerate}
    \setlength\itemsep{0.5em}
    \item Spread of the hoax:  When a susceptible agent $i$ interacts with a believer at time step \(t\), this agent may adopt the hoax and become a believer with a probability given by:
    \begin{equation*}
        f_i(t)=\beta \frac{n_i^B(t) (1+\alpha)}{n_i^B(t) (1+\alpha)+n_i^F(t) (1-\alpha)},
    \end{equation*}
    where \(n_i^B(t)\) and \(n_i^F(t)\) are the counts of the $i$-th agent's neighbors in the believer and fact-checker states, respectively.
    
    \item Fact-checking and debunking: Believers may fact-check the hoax and become fact-checkers with probability $p_{verify}$. Likewise, susceptible agents can also transition directly to a fact-checker state, with the following probability: 
    \begin{equation*}
    g_i(t)=\beta \frac{n_i^F(t) (1-\alpha)}{n_i^B(t) (1+\alpha)+n_i^F(t) (1-\alpha)}.
    \end{equation*}
     Once in the fact-checker state, individuals stop spreading the hoax and instead disseminate debunking information.
     
    \item Forgetting: Both believers and fact-checkers can forget their current belief state over time and revert to the susceptible state. This process is governed by a single probability $p_{forget}$ and applies to both believers and fact-checkers.
\end{enumerate}

Let $s_i(t)$ represent the state of the $i$th agent at time $t$, and define the state indicator function for $X \in \{B, F, S\}$ as $s_X^i(t) = \delta(s_i(t), X)$. The triple $p_i(t) = (p_B^i(t), p_F^i(t), p_S^i(t))$ describes the probability that node \( i \) is in each of the three states at time \( t \).




\subsection*{Network Generation}

We use the SBFC model to understand how network density influences the spread of misinformation in both segregated and unsegregated structures. Our analysis considers networks composed of two groups of different sizes---a majority group and a minority group---while also accounting for the densities of inter- and intra-group ties.

To generate segregated networks, we use the edge probability matrix to independently control the probability of intra- and inter-group links~\cite{oliveira2022group,karimi2023inadequacy}.
The edge probability matrix \( H \) is defined as follows:
\begin{equation}
H = \left[ \begin{array}{cc}
h_{00} & h_{01} \\
h_{01} & h_{11}
\end{array} \right],
\end{equation}
where $h_{00}$ and $h_{11}$ are the probabilities of intra-group edges, whereas $h_{01}$ is the probability of creating an inter-group edge.
In our work, we denote $f_{0}$ as the proportion of nodes belonging to the minority group.

\section*{Results}
We investigate how network density influences the spread of misinformation in both segregated and unsegregated networks. To model unsegregated networks, we use the Erdős-Rényi model, whereas we construct segregated networks by tuning $h_{00}$, $h_{11}$, and $h_{01}$.
This setup allows us to examine the effects of group size and connection patterns on misinformation spreading.

\subsection*{Unsegregated Networks}
\begin{figure}[b!]
     \centering
     \begin{subfigure}[b]{0.46\textwidth}
         \centering
         \includegraphics[width=\textwidth]{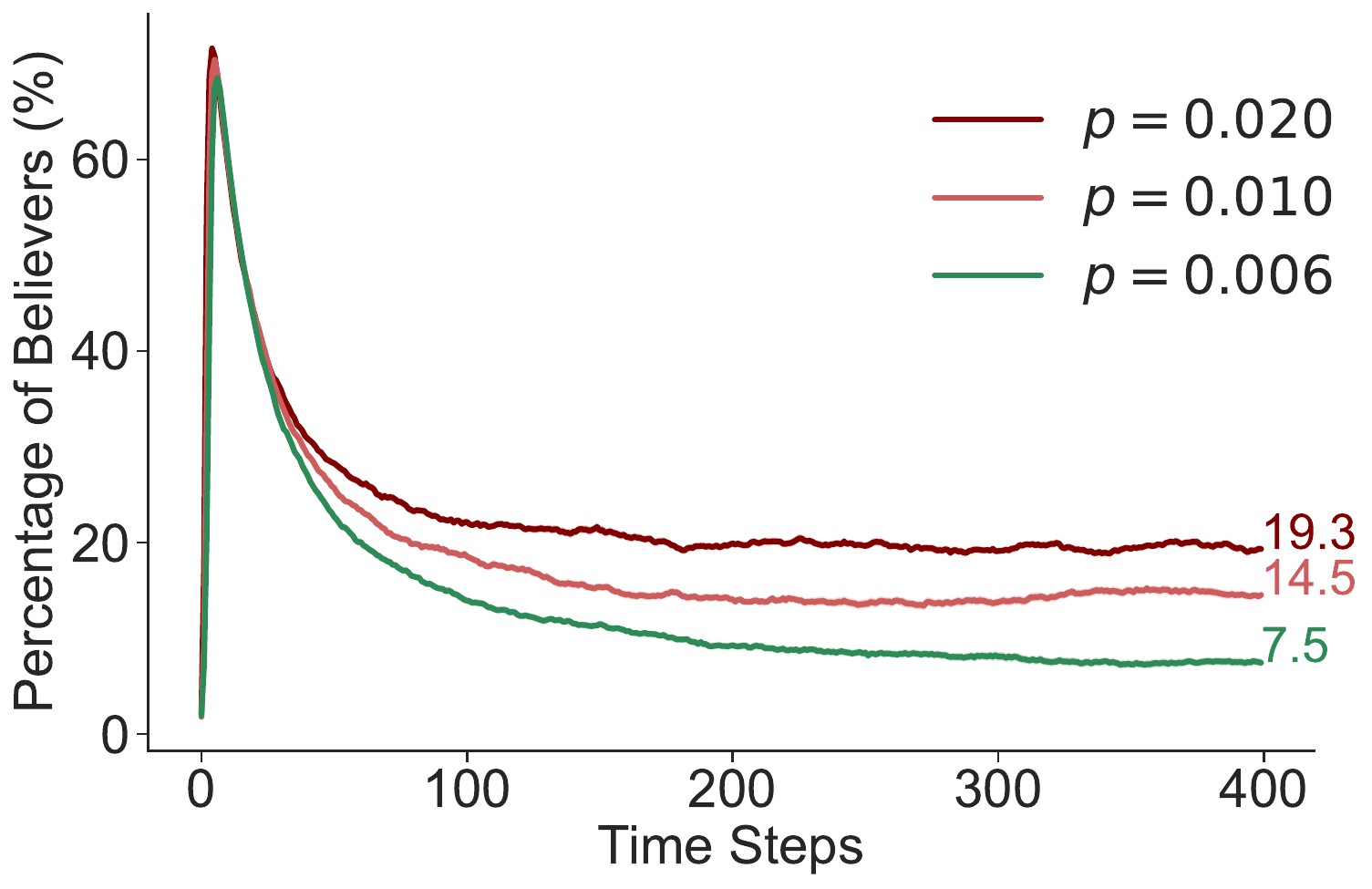}
         \caption{}
         \label{fig:time}
     \end{subfigure}
     \hfill
     \begin{subfigure}[b]{0.51\textwidth}
         \centering
         \includegraphics[width=\textwidth]{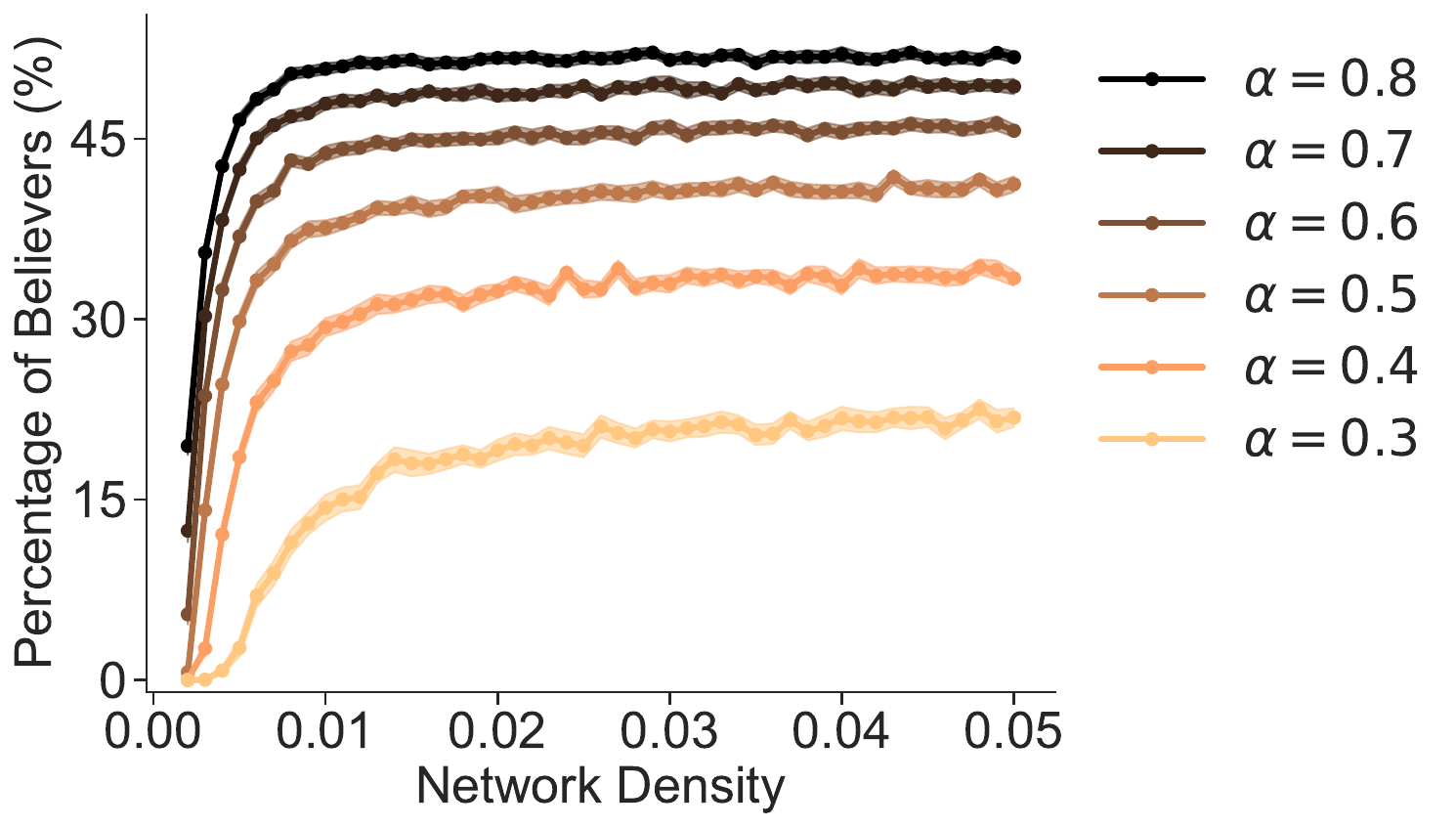}
         \caption{}
         \label{fig:density_alpha}
     \end{subfigure}
        \caption{\textbf{(a)} Percentage of Believers over time ($\alpha$ = 0.3).
        \textbf{(b)} Effect of Network Density on Percentage of Believers (\( n = 1000 \),  \( \text{iterations} = 50 \), \( p_{\text{verify}} = 0.05 \), \( p_{\text{forget}} = 0.1 \), \( \beta = 0.5 \)).}
     
\end{figure}

In unsegregated networks, our results show that denser networks lead to more believers. Fig.~\ref{fig:time} shows the impact of network density on the percentage of believers with the same level of gullibility ($\alpha$ = 0.3), where the parameter $p$ represents the probability of any two nodes in an Erdős-Rényi network. As $p$ increases (i.e., a higher network density), the percentage of individuals endorsing the misinformation also increases. 


We find that this result holds regardless of gullibility level. Fig.~\ref{fig:density_alpha} shows the final percentages of believers in a network comprising 1000 nodes, with varying values of the parameter $p$ and gullibility $\alpha$. Each line in the plot represents a different level of gullibility, indicating how belief formation is influenced by both network density and individual susceptibility. Across all lines representing different levels of gullibility, an increase in $p$ corresponds to a higher percentage of believers. 
%
This trend underscores the role of network density in facilitating the spread of misinformation, highlighting the importance of understanding network structure in combating false beliefs.

\subsection*{Segregated Networks}
        \begin{figure}[b!]
            \centering
            \includegraphics[width=1.\linewidth]{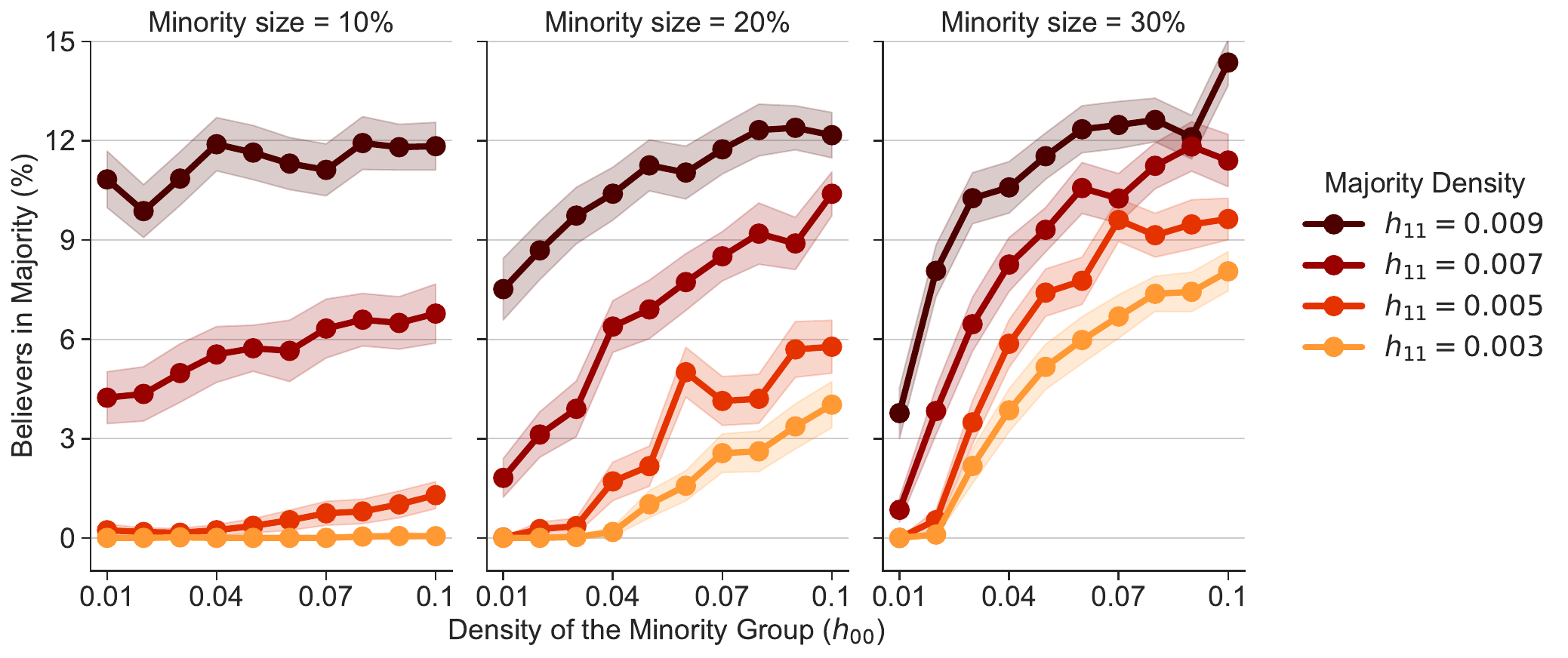}
            \caption{Percentage of Believers in the Majority for Different Minority Network Densities (\( n = 1000 \), \( h_{01} = 0.002 \), \( \text{steps} = 1000 \), \( \text{iterations} = 100 \), \( p_{\text{verify}} = 0.05 \), \( p_{\text{forget}} = 0.1 \), \( \beta = 0.5 \), \( \alpha = 0.3 \))}
            \label{fig:minority_effect}
        \end{figure}

In segregated networks, we find that increasing the density of the minority group not only increases the percentage of believers within that group but also affects the majority group, especially as the minority size grows. We examine the average percentage of believer members in the majority group across varying minority group sizes, ranging from 10\% to 30\% of the total population (see Fig.~\ref{fig:minority_effect}). 
%
%
 Consistent with the results of unsegregated networks, we find that an increase in the density of the majority group (i.e., $h_{11}$) yields a higher percentage of believers within this group.

Remarkably, increasing the density within the minority group (i.e., $h_{00}$) raises the percentage of believers in the majority group, with this effect intensifying as the minority group size increases. For example, with $h_{11} = 0.007$ and a minority group comprising 20\% of the population, increasing the density of the minority group results in a five-fold increase in believers within the majority group, from around 2\% to 10\%.

\section*{Discussion}

Our work demonstrates the critical role of network density in the spread of misinformation. Denser networks, whether unsegregated or segregated, foster a greater prevalence of false beliefs. In segregated networks, a dense minority group not only increases the percentage believers within itself but also amplifies this effect in the majority group. Here, the minority refers to a segregated group that shares the same characteristics as the majority group, except for having greater network density. These insights highlight the need to consider network structure when developing strategies to combat misinformation.

Our results indicate that targeting dense communities could be an effective strategy for combating the spread of misinformation. Such communities are not only highly suitable for the diffusion of misinformation but also amplify the percentage of believers in hoax news within the broader population. By targeting these dense clusters, interventions can be more effective in reducing the overall proliferation of misinformation.

Furthermore, our findings serve as a caution even for skeptical individuals with low gullibility: they may find themselves trapped in dense communities, inadvertently playing a role in spreading misinformation. Despite their skepticism, the high frequency of interactions within these networks increases the likelihood that they encounter and share false information.

In summary, our study sheds light on how network effects influence the spread of misinformation, providing a foundation for understanding its underlying mechanisms. Future research should explore other network configurations and real-world scenarios to guide the design of interventions aimed at building societal resilience against misinformation.

\bibliography{main_arxiv}
\bibliographystyle{naturemag}

\end{document}